# Methane as a dominant absorber in the habitable-zone sub-Neptune K2-18 b

Bruno Bézard[1*], Benjamin Charnay[1], Doriann Blain[1, 2]

[1]LESIA, Observatoire de Paris, Université PSL, CNRS, Sorbonne Université, Université Paris Cité, Meudon, France
[2]Presently at: Max Planck Institute for Astronomy, Königstuhl 17, D-69117 Heidelberg, Germany

[*]e-mail: bruno.bezard@obspm.fr

The transiting exoplanet K2-18 b, discovered in 2015 by the Kepler spacecraft[1], orbits a M3-dwarf (effective temperature $T_*$ = 3457 ± 39 K, radius $R_*$ = 0.411 ± 0.038 time the solar radius)[2] at a distance of 0.143 ± 0.006 AU (ref. [2]). These characteristics imply that K2-18b receives essentially the same insolation as the Earth does from the Sun. With a mass $M_p$ = 8.63 ± 1.35 Earth masses[3] and a radius $R_p$ = 2.61 ± 0.087 Earth radii[4], K2-18 b is considered as a super-Earth or sub-Neptune. Following the transits of K2-18 b observed by Kepler/K2 in the visible domain, planetary transits were observed with the Spitzer Space Telescope at wavelengths of 4.5 μm (ref. [5]) and 3.5 μm (ref. [4]). Nine transits were also observed with the Wide Field Camera 3 aboard the Hubble Space Telescope (HST)[4] and analysed by Tsiaras et al.[6]. The data that cover the range 1.12-1.63 μm clearly show an increase in transit depth at 1.4 μm (Fig. 1, Extended Data Fig. 1), coincident with a vibrational band of water vapour ($H_2O$). Using a model atmosphere and a retrieval algorithm, Tsiaras et al. concluded to the clear detection of water vapour, determined that other gases such as carbon monoxide (CO), carbon dioxide ($CO_2$), ammonia ($NH_3$) or methane ($CH_4$) were not present in measurable quantities, and concluded that a significant fraction of the atmosphere is still made of hydrogen and helium. Another simultaneous study corroborates this interpretation[4]. However, for such a mini Neptune-type planet, one would expect significant amounts of compounds beyond $H_2O$, such as $CH_4$ or $NH_3$. In this Matters Arising we argue that the reported absorption is most likely due to methane.

We modelled the composition of K2-18 b using the Exo-REM radiative-equilibrium model[7-9] adapted for irradiated giant exoplanets[10], and we calculated the expected transit spectra. Beyond $H_2$/He, Exo-REM incorporates 12 gaseous absorbers including all potential ones in the HST transit data, i.e. $H_2O$, CO, $CO_2$, $CH_4$, $NH_3$ and $H_2S$, but no cloud opacity. Mixing ratio profiles are calculated either from thermochemical equilibrium or allowing for some non-equilibrium chemistry between C-, O- and N-bearing compounds[11]. We investigated a range of metallicity of (1-1000), defined here as the ratio of the abundance of elements heavier than hydrogen or helium (relative to hydrogen) to that in the Sun. For the eddy mixing coefficient ($K_{zz}$), a parameter that controls the transport-induced quenching of chemical equilibria, we investigated a range of $10^6$-$10^{10}$ cm$^2$ s$^{-1}$.

Transmission spectra were calculated and compared with the HST transit depths[6] along with the K2 and Spitzer transit depths[4]. The radius of the planet at a reference pressure level was set, for each model, to the value that minimizes the residuals with the data in a least square sense ($\chi^2$). The goodness of the fit was then estimated from this minimum value of $\chi^2$. With 19 degrees of freedom (17 HST + 1 K2 + 2 Spitzer data points – 1 free parameter [radius]), any value exceeding 21.4 would indicate that the model does not reproduce the data at the 1$\sigma$ (68%) confidence level. We obtained self-consistent models that agree with the data for metallicities of 70-800 and any value of $K_{zz}$. For these models, the temperature in the upper atmosphere probed by the HST data (~0.001-30 mbar) is in the range 240-300 K, the $CH_4$ mole fraction is 0.03-0.10, and that of $H_2O$ 0.05-0.11 (e.g. Extended Data Fig. 2). The calculated transit absorption spectra exhibit a marked maximum around 1.4 µm, in agreement with the HST data. However, this maximum is predominantly due to $CH_4$ absorption rather than to $H_2O$ as previously concluded (Fig. 1). We would need to reduce the $CH_4$ mole fraction by an order of magnitude to reach similar absorption levels for $H_2O$ and $CH_4$, even though the $H_2O$ band ($\nu_1+\nu_3$ and $2\nu_1$) is intrinsically ~20 times stronger than the $CH_4$ band (isocad) and the abundances of the two species differ by at most 60%. The difference in absorptivity is due to the structure of the molecules. The infrared spectrum of $CH_4$ is much more congested than that of $H_2O$ because of a larger number of vibrational degrees of freedom, and couplings among overlapping bands[12]. At the temperatures (240-300 K) and low pressures probed by the transit spectra, pressure broadening is weak, the strong lines are saturated, and absorption occurs mostly through the much more numerous weak lines.

While our best-fitting model (metallicity of 180, $K_{zz}$ = $10^8$ cm$^2$ s$^{-1}$) including all absorbers agrees with the data at the 1$\sigma$ confidence ($\chi^2$ = 17.3), spectra in which only $CH_4$ absorption or only $H_2O$ absorption is considered yield equally good fits ($\chi^2$ = 18.4 and 17.7 respectively) (Fig. 1). Therefore, we agree with Tsiaras et al.[6] that $H_2O$ can provide the observed absorption at 1.4 µm but we disagree with the assertion that these HST data provide unambiguous evidence for its presence. In contrast, we find that $CH_4$ is by far the dominant absorber at this wavelength, assuming a Neptune-type composition with a moderately large metallicity, a case that was not considered in the three scenarios they investigated. The 1.4-µm band alone is thus more diagnostic of the presence of methane than of water vapour for relatively cold giant planets such as K2-18b. Absorption from water vapour would dominate over methane only or if more than 90% of the carbon is sequestrated in CO rather than in $CH_4$, contrary to expectations from chemical models, or if the planet's C/O ratio is 8-26 times lower than the protosolar value (0.55), a possibility that we regard as unlikely.

We have investigated up to which atmospheric temperature $CH_4$ absorption dominates over $H_2O$ absorption at 1.4 µm in the transit spectra of sub-Neptunes. First, we kept the abundance profiles from our best-fitting atmospheric model (Extended Data Fig. 2), but assumed an isothermal atmosphere in the region probed by the transit spectra and varied its temperature ($T_{atm}$). As $T_{atm}$ increases, more and more energy levels are populated and absorption occurs through an increasing number of transitions for both molecules. Methane then gradually loses its spectroscopic specificity at low temperatures and, for $T_{atm}$ > 1000 K, absorption by the stronger water band prevails over that of methane (Fig. 2). In a second step, we considered self-consistent models in which we fixed the metallicity to 180 and varied the planet's effective temperature ($T_{eff}$) by modifying its distance to the star. In this case, $H_2O$ absorption prevails over $CH_4$ absorption for $T_{eff}$ > 600 K (Fig. 2). While this estimate relies on known chemical processes, a caveat is the depletion of methane observed in GJ3470 b, a sub-Neptune with a low metallicity and $T_{eff}$ ~ 600 K, which suggests a possible inhibition of the CO to $CH_4$ conversion at deep hot levels[13].

We have shown that the 1.4-µm region alone cannot be diagnostic of the presence of water vapour for the cool planet K2-18b and even for mini-Neptunes having an effective temperature up to 600 K, the absorption being likely mostly due to methane. A confusion arises from the similarity of the spectra of the two molecules from 1.10 to 1.55 µm, as discussed in a previous study[14]. Data from other spectral ranges, particularly in the interval 1.6-3.7 µm, would allow us to clearly discriminate between $H_2O$ and $CH_4$ absorption, and also, in principle, determine the abundance ratio of the two species (Extended Data Fig. 1). Such a measurement would be very important to understand the internal structure of K2-18b and the location of a possible liquid water ocean[15].

## Methods

**Self-consistent atmosphere models.** We modelled the atmosphere of K2-18 b with Exo-REM, a one-dimensional radiative-equilibrium model for giant planets[7-9], recently adapted for irradiated planets[10]. Exo-REM solves for the planet-average temperature and chemical composition profiles (Extended Data Fig. 2), based on the conservation of the total (radiative + convective) flux and assuming either chemical equilibrium or allowing for quenching of the equilibria of CO-$CH_4$, CO-$CO_2$, and $NH_3$-$N_2$. In the latter case, the $CH_4$, $CO_2$ and $N_2$ abundances are held constant above their respective quench levels, which are defined by the equality of interconversion chemical time and atmospheric mixing time (parametrized by the eddy mixing coefficient $K_{zz}$). The chemical timescales are calculated from simple functional forms that span a range of warm extrasolar giant planets[11]. No photochemical processes are included. Exo-REM includes opacity from the collision-induced absorption of $H_2$-$H_2$, $H_2$-He and $H_2O$-$H_2O$ pairs and from molecular bands of $H_2O$, $CH_4$, CO, $CO_2$, $NH_3$, $H_2S$, $PH_3$, TiO, VO, FeH, Na and K. Exo-REM makes use of k-correlated absorption coefficients calculated on a pressure-temperature grid using the HITEMP ($H_2O$, CO, $CO_2$), TheoReTS ($CH_4$), NIST (Na, K) and ExoMol (all other species) databases (see ref. [10] for details and references). We have checked that using the more complete POKAZATEL[16] line list for $H_2O$ does not noticeably produce more absorption in the calculated transit spectra. While self-consistent cloud models may now be included in Exo-REM[9], here we chose to consider only cloud-free atmospheres. Simulations of the structure of K2-18 b with a General Circulation Model (GCM)[17] show that the $H_2O$ cloud cover is strongly affected by dynamics, being quite inhomogeneous in latitude and longitude, and in some cases, varying with time. We thus believe that it cannot be reliably treated in a one-dimensional radiative-convective equilibrium model. Preliminary GCM calculations show that, for metallicities from 100 to 300, the fractional cloud cover at the limbs is much less than unity and its effect on the transit spectrum in the HST/WFC3 range is very weak. We finally note that the presence of clouds/hazes was not conclusively detected in a recent study of the atmosphere of K2-18 b (ref. [15]) in contrast to previous suggestions[4,6].

We modelled the stellar flux using a spectrum of GJ 176, a M2.5-star with an effective temperature of 3416 K (ref. [18]) and similar to K2-18 in other stellar properties. We extracted the GJ 176 spectrum from the MUSCLES database (version 2.2)[19] (https://archive.stsci.edu/prepds/muscles/), extrapolated it beyond 5.5 µm with a Planck function at 3200 K to ensure continuity at this wavelength, and rescaled it by the ratio of the Planck functions at 3460 K and 3416 K to obtain a stellar spectrum at the 3457-K effective temperature of K2-18 (ref. [5]). We calculated the planet's irradiation using a star radius of 0.4445 times the Sun's radius, a star-planet distance of 0.1591 AU (ref. [4]) and a geometric factor of 0.25, to account for planet-averaged irradiation conditions compared to normal incidence. We added an internal heat source equivalent to an internal temperature $T_{int}$ = 60 K, comparable to that of Neptune ($T_{int} \approx$ 50 K), and typical of that expected for a planet with K2-18 b's mass and radius, older than a few Gyr (ref. [20]). The acceleration of gravity was calculated for a planet's mass of 8.63 Earth masses (ref. [4]) and a 1-bar radius of 16,430 km. The spectral flux (stellar and planetary thermal emission) was modelled from 30 to 30,010 cm$^{-1}$ with $k$-correlated coefficients defined over 20-cm$^{-1}$ intervals. We ran Exo-REM for metallicities ($M$) varying from 1 to 1000, as compared with the present solar-system elemental composition[21]. This range encompasses that found in the Solar System (from ~4 for Jupiter to about 80-200 for Neptune) and that expected for a 10 Earth-mass giant planet[22] (20-400). The He/H ratio was fixed at 0.0839 (ref. [21]). We investigated a range for the eddy mixing coefficient ($K_{zz}$), that parametrizes the strength of the vertical transport and thus the efficiency of the chemical quenching, from $10^6$ to $10^{10}$ cm$^2$ s$^{-1}$. We also tested models at local thermochemical equilibrium. The atmospheric structure was simulated over a grid of 64 levels with pressures ranging from 200 bars to $10^{-5}$ mbar, sampled uniformly in log space.

**Spectral calculations and comparison with data.** We calculated transit spectra at 20-cm$^{-1}$ resolution by radially integrating the slant path transmittance over the Exo-REM atmospheric grid and using 64 lines-of-sight tangent to the pressure levels of the grid. We compared these synthetic spectra with the dataset consisting of the HST/WFC3 data around 1.4 µm, as reduced by Tsiaras et al.[6], the K2 broad-band visible measurement[4] and two Spitzer/IRAC measurements[4] at 3.5 and 4.5 μm (Extended Data Fig. 1). We then averaged the synthetic spectra over the bandwidths of these measurements and calculated the chi-squared test statistic ($\chi^2$) equal to the sum of normalized squared deviations between observations and model (20 data points). In this comparison, a free parameter is the radius at some reference pressure level in our simulations (e.g. 1 bar) and, for each model, we shifted our calculated radius by the value that minimizes the $\chi^2$. This minimum value $\chi_{min}^2$ provides an estimate of the goodness of the fit. Considering 20 - 1 = 19 degrees of freedom, models yielding $\chi^2 < 21.4$ are consistent with the data at the 1σ confidence level ($\chi^2 < 30.5$ at the 2σ confidence level).

We found that models with $M$ between 70 and 800 and any value of $K_{zz}$ in the range that we investigated allow us to reproduce the observations within the 1σ confidence level. In the upper atmosphere, the $H_2O$ mole fraction is uniform and varies with $M$ and $K_{zz}$ from 0.051 to 0.112, the $CH_4$ mole fraction from 0.033 to 0.098, the CO mole fraction from $0.15 \times 10^{-2}$ to 0.17, and the $NH_3$ mole fraction from $0.58 \times 10^{-3}$ to $0.85 \times 10^{-3}$. We also compared our simulations with a dataset incorporating the HST data reduced by Benneke et al.[4] in place of those from Tsiaras et al.[6]. In this case, only models with $M$ of 150-200 agree with the data at the 1σ level. We also found that the fit of the HST data is clearly superior when only $H_2O$ absorption is included ($\chi_{min}^2$ = 12.8), in agreement with previous investigations[4,15] (Extended data Fig. 3). This $H_2O$-only model is thus overfitting the HST data reduced by Benneke et al.[4], while the complete model with all absorbers ($\chi_{min}^2$ = 20.9) is still at the 1σ confidence level, and thus compatible with this dataset. In any of these cases, $CH_4$ absorption dominates over $H_2O$ absorption at 1.4 μm. We ran tests in which we decreased the $CH_4$ mole fraction in our model, otherwise keeping the same atmospheric structure and $H_2O$ mole fraction, until the $CH_4$-only spectrum yields the same radius as a $H_2O$-only spectrum. We conclude from this analysis that we would need to reduce the $CH_4$ mole fraction by a factor of 9-16 (for metallicities of 70-800) to reduce $CH_4$ absorption at the level of that of $H_2O$.

In models with $M$ below 180, water vapour does not condense out and water clouds are therefore not expected. At higher metallicities, water condensation is predicted to occur between levels (7 mbar, 250 K) for $M$ = 190 and (18 mbar, 264 K) for $M$ = 800. Because the condensation region does not extend above the 6-mbar region, cloud opacity is expected to have only a moderate effect on the transit spectrum.

**Comparison with an atmospheric retrieval approach.** To check our radiative transfer model, we simulated the transit spectrum of K2-18 b for the same atmospheric composition as shown in Extended Data Fig. 2, using the forward model of TauREx 3 (ref. [23] and used by Tsiaras et al.[6]) accessible online (https://taurex3-public.readthedocs.io/en/latest/). As with Exo-REM, we find that $CH_4$ absorption dominates at 1.4 micron despite the higher abundance of $H_2O$ (Extended data Fig. 4). We then performed a retrieval using TauREx 3 for the HST data as reduced by Tsiaras et al.[6]. We used as free parameters: planetary radius, temperature, abundances of $H_2O$, $CH_4$ and $N_2$, and cloud-top pressure. The volume mixing ratios of $H_2O$, $CH_4$ and $N_2$ are allowed to vary from $10^{-10}$ to 0.5. The posterior distributions of log($H_2O$) and log($CH_4$) are very broad with relative flat maxima in the range [-3.5:-1] and [-4:-0.3] respectively (Extended Data Fig. 5). A large fraction of the solutions has a $CH_4$ mixing ratio larger than a tenth of the $H_2O$ one, corresponding to a $CH_4$-rich atmosphere whose absorption at 1.4 micron is dominated by $CH_4$. In particular the log($H_2O$) and log($CH_4$) of our best fit Exo-REM model having a metallicity of 180 times solar (-0.96 and -1.2 respectively) are located in a dense region of the parameter space (Extended Data Fig. 5). A Neptune-like planet, with significant amounts of methane, was not one of the three scenarios tested by Tsiaras et al.[6], which likely explains why these authors found evidence for $H_2O$ and excluded $CH_4$ as a significant source of opacity at 1.4 µm.

## Data availability
The observational data that support the findings of this study are included in refs. [4,6]. The model data that support the plots within this paper are available from the corresponding author upon request.

## Code availability
The Exo-REM source code and the input files used in this study are available from the corresponding author upon reasonable request. The current version of Exo-REM is available through the Observatoire de Paris GitHub website https://gitlab.obspm.fr/Exoplanet-Atmospheres-LESIA/exorem.

## Acknowledgements
D.B. acknowledges financial support from the ANR project « e-PYTHEAS » (ANR-16-CE31-0005-01). This work received funding from the Programme National de Planétologie (PNP) of CNRS/INSU, co-funded by CNES.

## Author contributions
B.B. initiated the development of the Exo-REM model, performed the analysis, and wrote the paper with the help of B.C. and D.B. B.C. contributed to the development of the Exo-REM model, compared Exo-REM and TauREx outputs, and wrote the relevant section. D.B. calculated the tables of absorption *k*-coefficients and adapted the Exo-REM model to irradiated planets, with contributions from B.B and B.C. All authors contributed to the interpretation of the results and commented on the manuscript at all stages.

## Competing interests

The authors declare no competing interests.

# Figures

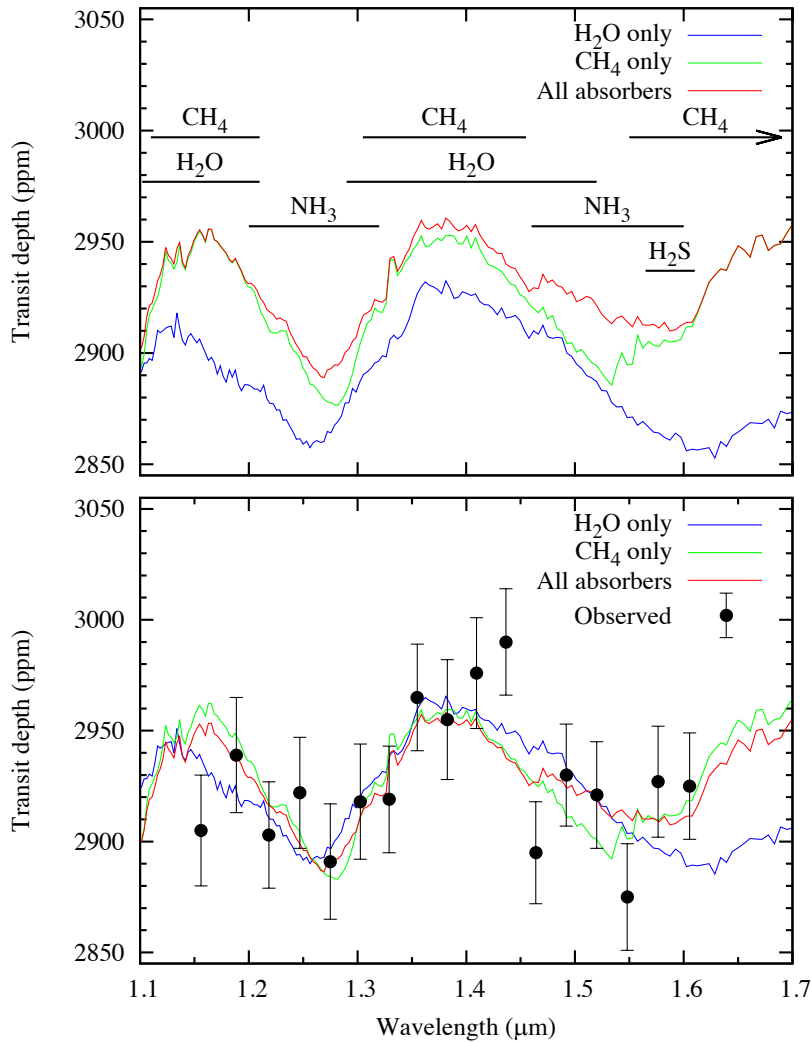

**Fig. 1 | Transmission spectra of K2-18b calculated for different atmospheric compositions.** Top: Transmission spectrum calculated from our self-consistent model Exo-REM for a metallicity of 180 and an eddy mixing coefficient of $10^8$ cm$^2$ s$^{-1}$ (red). Spectra using the same atmospheric model but including absorption from H$_2$O only (blue) or from CH$_4$ only (green) are also shown. The dominant molecular absorbers are labelled. At 1.4 μm, a spectrum including only CH$_4$ absorption is only 5 ppm below the nominal one (i.e. including all absorbers) whereas the difference amounts to 30 ppm if H$_2$O absorption alone is considered. Bottom: Same synthetic spectra compared with HST data[6] after adjusting the planet's radius in each case to minimize the residuals.

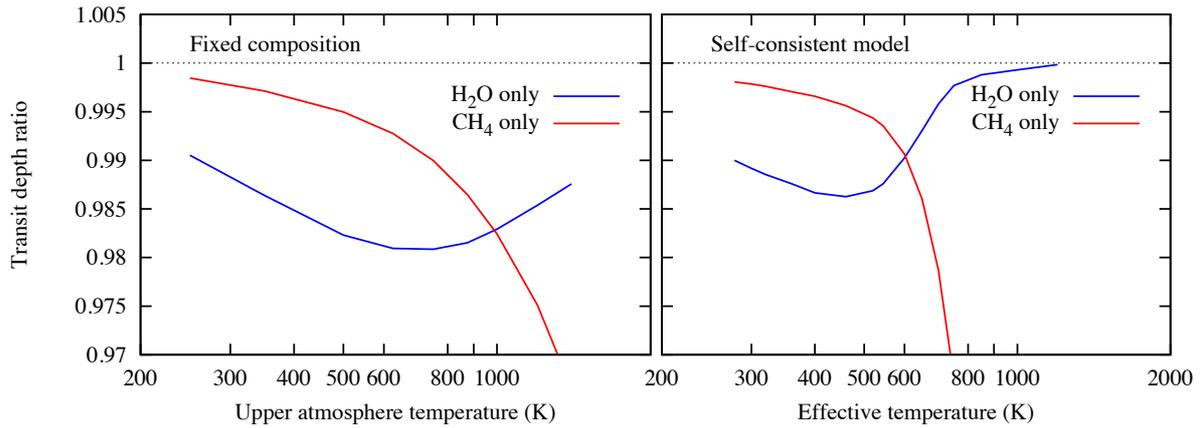

**Fig. 2 | Relative contributions of H₂O and CH₄ to the transit depth.** Shown is the ratio of the transit depth averaged over 1.335-1.415 µm assuming only H$_2$O (blue) or CH$_4$ (red) absorption, to the transit depth including all absorbers. Left: composition is that of our best fitting model (having H$_2$O and CH$_4$ mole fractions of 0.109 and 0.068 respectively in the upper atmosphere) and only the temperature of the upper atmosphere is varied. Right: composition and temperature profiles are calculated from our self-consistent model Exo-REM model for a metallicity of 180 and an eddy mixing coefficient of $10^8$ cm$^2$ s$^{-1}$. In this case, H$_2$O absorption prevails over CH$_4$ absorption for effective temperatures ($T_{eff}$) above 600 K, a value lower than in the left panel because the abundance of CH$_4$ decreases with increasing $T_{eff}$, carbon getting preferentially bound in CO.

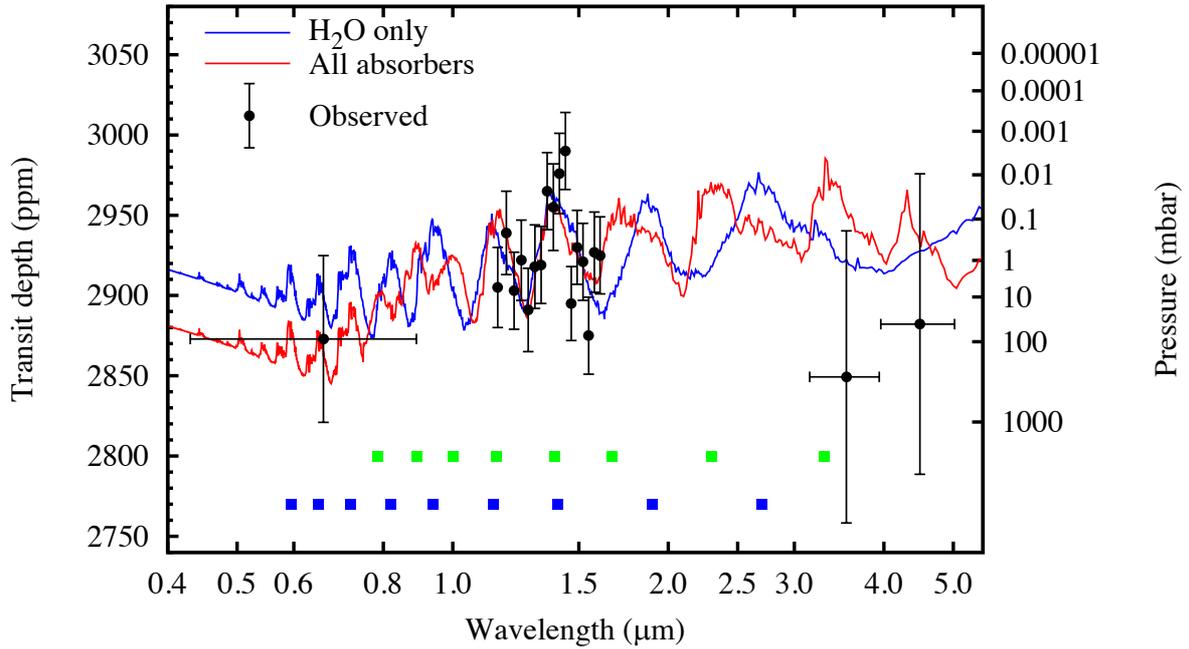

**Extended Data Figure 1 | Comparison of model transmission spectra with the observational data set.** The transmission spectrum of K2-18b calculated from our best-fitting self-consistent model (red) is compared with the HST/WFC3 data[6], the Kepler/K2 data point[4] and the Spitzer/IRAC measurements[4] at 3.5 and 4.5 μm. A model including only $H_2O$ absorption is shown in blue. For each model, the planet's radius has been adjusted to minimize the $\chi^2$ residuals. The location of the main absorption bands of $H_2O$ (blue squares) and $CH_4$ (green squares) are indicated. While both models are similar in the spectral range covered by the HST data, they vastly differ at longer wavelengths where the $CH_4$ bands no longer coincide with the strong $H_2O$ bands. The rightmost vertical axis represents the pressure levels for our best-fitting model having a metallicity of 180 (red).

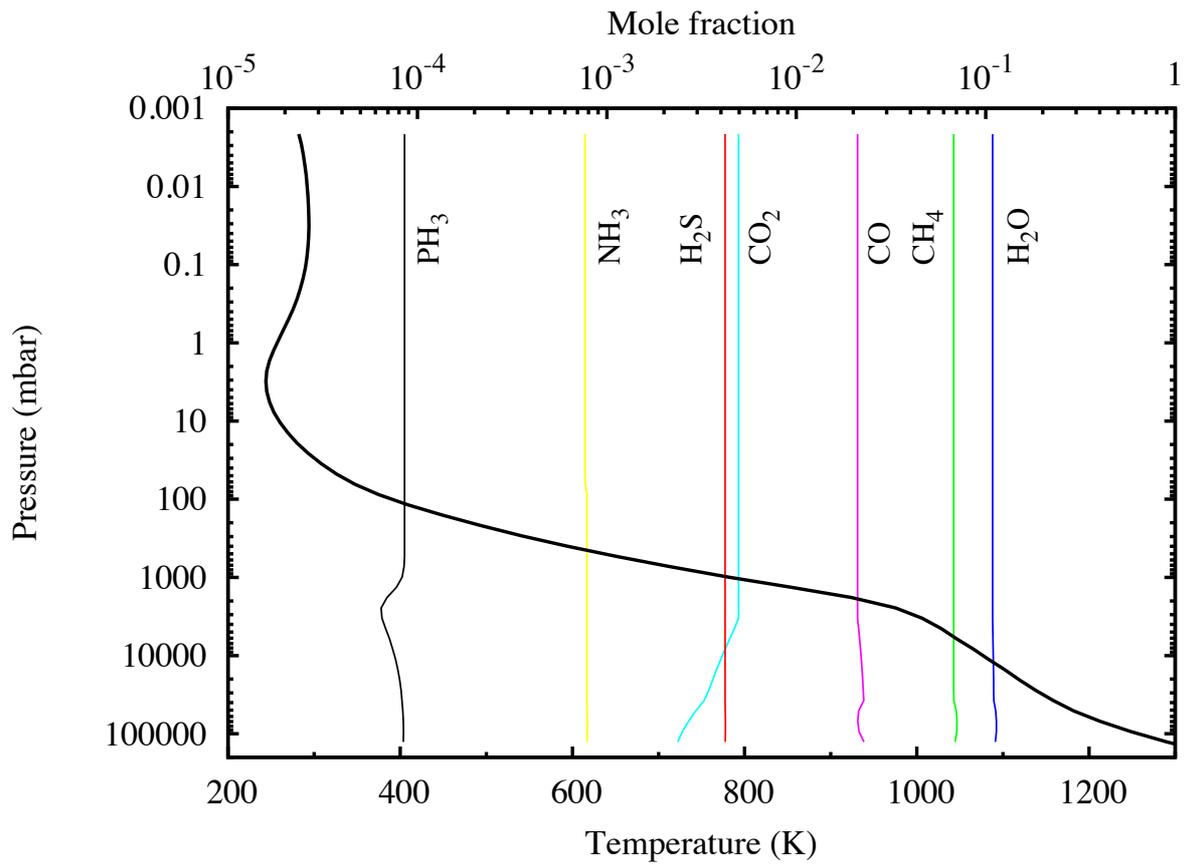

**Extended Data Figure 2 | Best-fit atmospheric model.** Temperature profile (thick black line) and abundance profiles of selected molecular absorbers (coloured lines) for our best-fit Exo-REM model. The metallicity is 180 and the eddy mixing coefficient is $10^8$ cm$^2$ s$^{-1}$. In this model, water vapour does not condense and water clouds are not expected to form. The mean molecular mass is 6.2 amu.

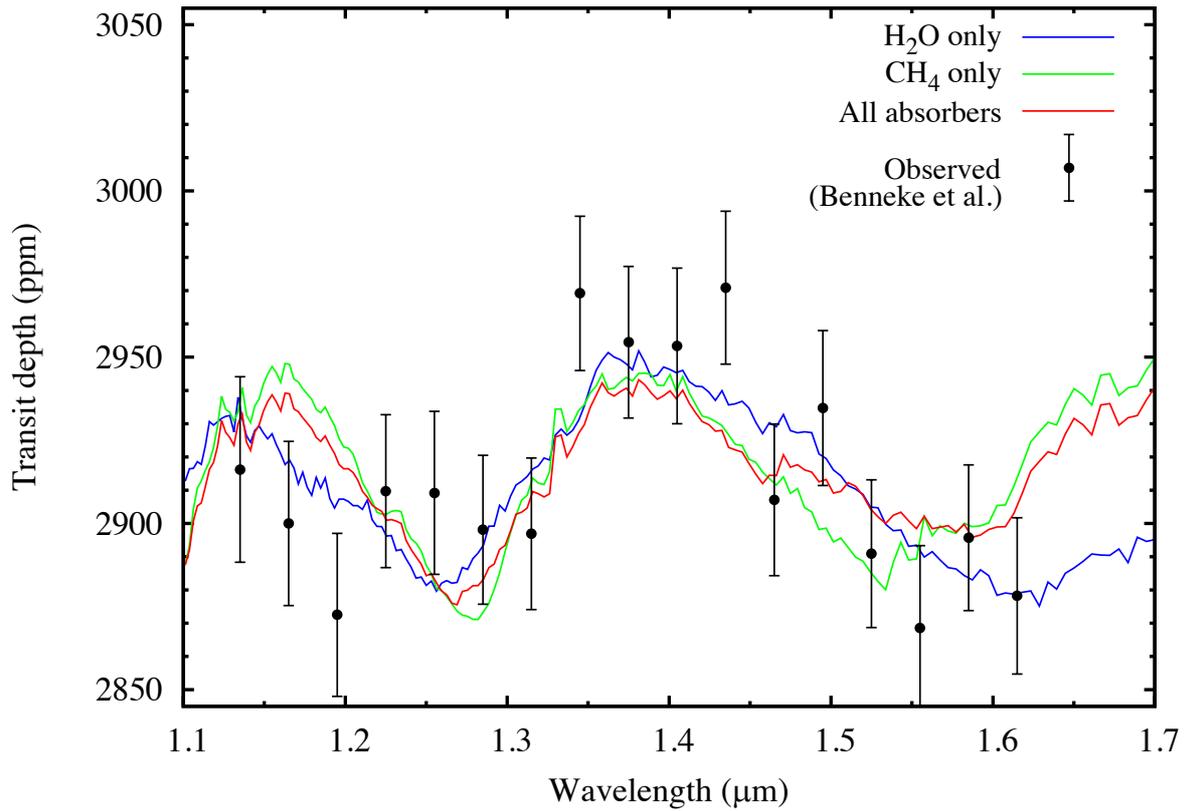

**Extended Data Figure 3. Transmission spectra of K2-18 b calculated for different atmospheric compositions.** Spectra calculated from our self-consistent model Exo-REM for a metallicity of 180 and an eddy mixing coefficient of $10^8$ cm$^2$ s$^{-1}$, are compared with HST data as reduced by Benneke et al.[4]. All absorbers are included in the spectral calculation shown in red. Spectra including absorption from $H_2O$ only (blue) and from $CH_4$ only (green) are also shown. The planet's radius has been adjusted in each case to minimize the residuals with the whole dataset[4].

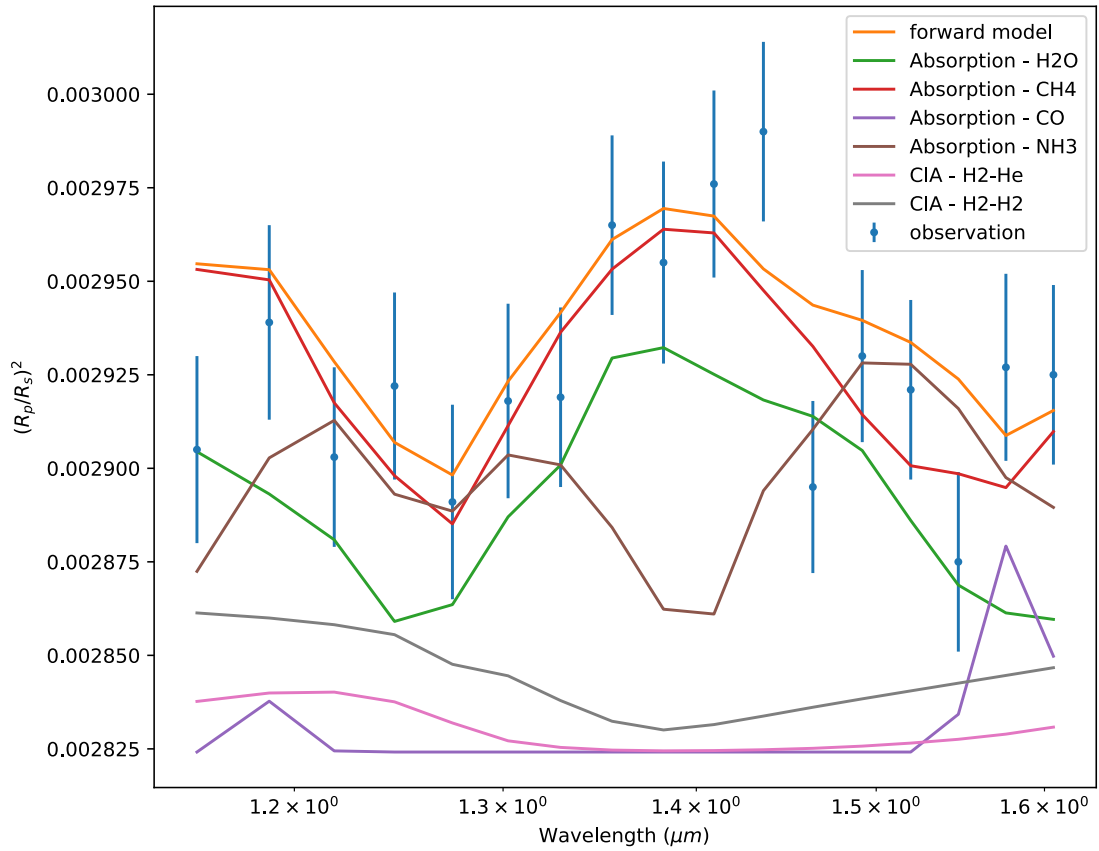

**Extended Data Figure 4 | Contributions of different absorbers to the transit depth.** Transmission spectra of K2-18 b calculated for the atmospheric composition shown in Extended Data Fig. 2 using the forward model of TauREx 3 (ref. [23]). The orange line corresponds to the spectrum with molecular absorption from $H_2O$, $CH_4$, CO, $NH_3$ and collision-induced absorption (CIA) from $H_2$-$H_2$ and $H_2$-He. Other coloured lines show the contribution of each molecule.

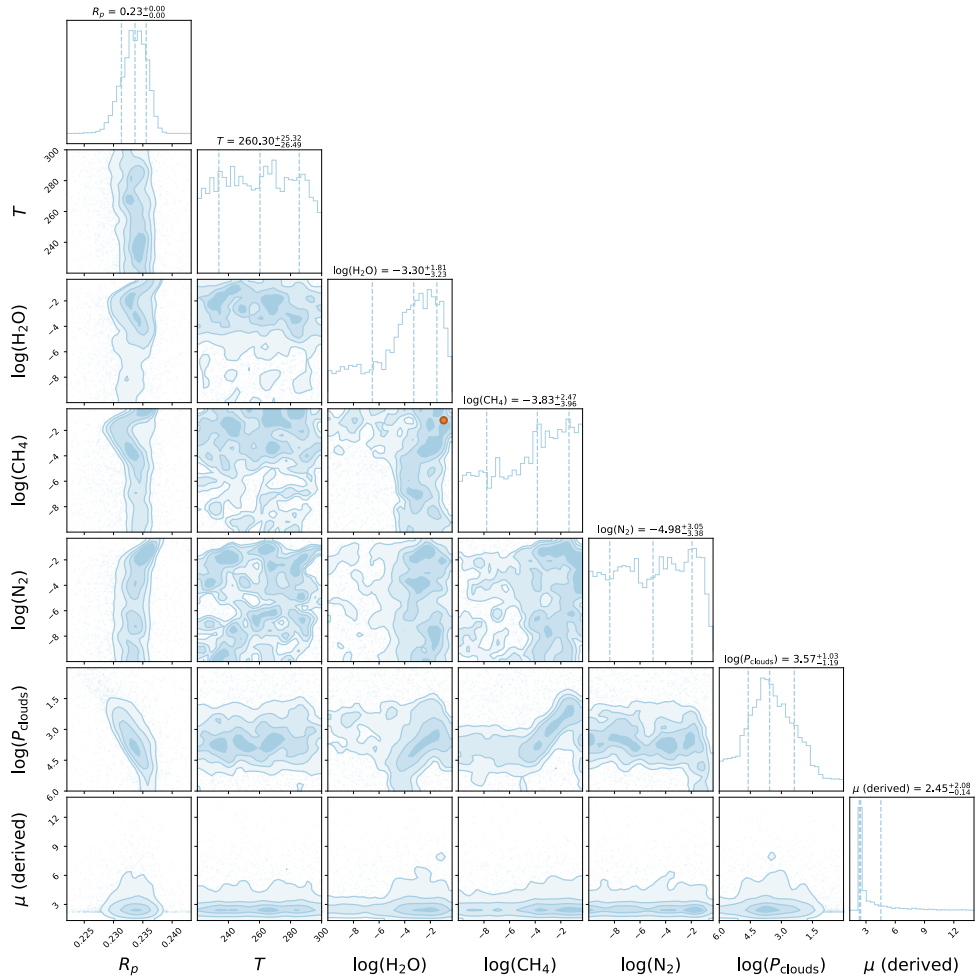

**Extended Data Figure 5 | Posterior distributions obtained from free retrievals.** The retrieval tool TauREx 3 (https://taurex3-public.readthedocs.io/en/latest/) was applied to the HST data reduced by Tsiaras et al.[6] with planetary radius (expressed in Jovian radius), temperature, log of $H_2O$, $CH_4$ and $N_2$ volume mixing ratios, and cloud-top pressure (Pa) as free parameters. The posterior distributions of the log($H_2O$) and log($CH_4$) are very broad with a denser region for the ranges [-3.5:-1] and [-4:-0.3] respectively. A large fraction of the parameter space corresponds to a $CH_4$-rich atmosphere whose absorption at 1.4 micron is dominated by $CH_4$. The parameters of our best fit Exo-REM model having a metallicity of 180 (Extended Data Fig. 2), shown as an orange dot in the log($CH_4$)-log($H_2O$) space, are globally consistent with the TauREx retrievals.